\documentclass[12pt]{article}
\usepackage{amsmath,amssymb,amsthm}
\usepackage{txfonts,bm,pifont}
\usepackage{graphicx}
\usepackage[pdfencoding=auto,bookmarks=true,bookmarksnumbered=true,colorlinks=true]{hyperref}
\usepackage{color}
\usepackage{float}
\usepackage{multirow,bigdelim}
\theoremstyle{plain}
\newtheorem{thm}{Theorem}[section]

\newtheorem{rem}[thm]{Remark}

\makeatletter
\newdimen\LENB \newdimen\LENW \newdimen\THI 
\newdimen\LENWH \newdimen\LENTOT \newcount\N 
\def\vbrknlnele#1#2#3{
  \LENB=#1pt \LENW=#2pt \THI=#3pt
  \LENWH=\LENW \divide\LENWH by 2
  \LENTOT=\LENB \advance\LENTOT by \LENW
  \vbox to \LENTOT{
    \vbox to \LENWH{}
    \nointerlineskip
    \vbox to \LENB{\hbox to \THI{\vrule width \THI height \LENB}}
    \nointerlineskip
    \vbox to \LENWH{}
  }}

\def\vbrknln#1{
  \N=#1
  \vcenter{
    \vbox{
      \loop\ifnum\N>0
        \vbox to 4pt{\vbrknlnele{2}{2}{0.1}}
        \nointerlineskip
        \advance\N by -1
      \repeat
  }}}

\def\hbrknlnele#1#2#3{
  \LENB=#1pt \LENW=#2pt \THI=#3pt
  \LENTOT=\LENB \advance\LENTOT by \LENW
  \vcenter{
    \vbox to \THI{
      \hbox to \LENTOT{
        \hfil
        \vrule width \LENB height \THI
        \hfil}
  }}}

\def\hblele{\hbrknlnele{2}{2.2}{0.1}}

\def\hblfil{\cleaders\hbox{$ \m@th \mkern1mu \hblele \mkern1mu
$}\hfill} 
\makeatother
\begin{document}
\begin{center}
\begin{large}
Log-aesthetic Curves as Similarity Geometric Analogue of Euler's Elasticae\\[5mm]
\end{large}
\begin{normalsize}
Jun-ichi {\sc Inoguchi}\\
Institute of Mathematics, University of Tsukuba\\
Tsukuba 305-8571, Japan\\
e-mail: inoguchi@math.tsukuba.ac.jp\\[2mm]
Kenji {\sc Kajiwara}\\
Institute of Mathematics for Industry, Kyushu University\\
744 Motooka, Fukuoka 819-0395, Japan\\
e-mail: kaji@imi.kyushu-u.ac.jp\\[2mm]
Kenjiro T. {\sc Miura}\\
Graduate School of Science and Technology,Shizuoka University\\
3-5-1 Johoku, Hamamatsu, Shizuoka, 432-8561, Japan\\
e-mail: miura.kenjiro@shizuoka.ac.jp\\[2mm]
Masayuki {\sc Sato}\\
Serio Corporation\\
2-21-10 Shinkawa, Chuo-ku, Tokyo 104-0033, Japan\\[2mm]
Wolfgang K. {\sc Schief}\\
School of Mathematics and Statistics, The University of New South Wales\\
Sydney, NSW 2052, Australia\\
e-mail: w.schief@unsw.edu.au\\[2mm]
Yasuhiro {\sc Shimizu}\\
UEL Corporation\\
1-1-1 Toyosu, Koto-ku, Tokyo 135-8560, Japan
\end{normalsize}
\end{center}
\begin{abstract}
In this paper we consider the log-aesthetic curves and their generalization which are used in
CAGD. We consider those curves under similarity geometry and characterize them as stationary
integrable flow on plane curves which is governed by the Burgers equation. We propose a variational
formulation of those curves whose Euler-Lagrange equation yields the stationary Burgers equation.
Our result suggests that the log-aesthetic curves and their generalization can be regarded as the
similarity geometric analogue of Euler's elasticae.
\end{abstract}
\section{Introduction}
In this paper we consider a class of plane curves in CAGD called the {\em log-aesthetic curves}
(LAC) and their generalization called the {\em quasi aesthetic curves} (qAC), and present their new
mathematical characterization by using the theory of integrable systems.
In 1744, Euler introduced the elastic energy of a plane curve which is the totality of squared
signed curvature, and studied a variational problem of plane curves in Euclidean geometry
\cite{Euler}. The solutions to this problem are called \emph{elasticae}.  On the other hand, a
standard isoperimetric deformation of arc length parametrized plane curves is governed by the {\em
modified Korteweg de Vries (mKdV) equation} for the signed curvature \cite{GP}, which is one of the
most important integrable equations. Then it was seen that elasticae are characterized as travelling
wave solutions to the mKdV equation.
In this paper, we consider LAC and qAC in the framework of \textit{similarity geometry}, and
characterize them as the similarity geometric analogue of elasticae.  We formulate them as
stationary curves with respect to the simplest integrable flow given by the {\em Burgers
equation}. We then give a variational formulation by introducing a functional (called the fairing
energy) for plane curves, from which we deduce the stationary Burgers equation as the Euler-Lagrange
equation. This gives a new mathematical characterization of LAC and qAC and demonstrates the
usefulness of Klein geometry and integrable systems as tools of CAGD.

\section{Elasticae and mKdV flows}
To clarify the mathematical background and motivations of this paper, we recall briefly here the
basic facts on the elasticae and the Goldstein-Petrich flows.  An arc length parametrized curve
$\gamma(s)$ in the Euclidean plane $\mathbb{E}^2$ is said to be an \textit{elastica} if it is a
critical point of the \textit{elastic energy}:
\begin{equation}
E(\gamma)=\int_{0}^{\ell}\frac{1}{2}
\kappa
(s)^2\,ds,
\end{equation}
through total length preserving variations.  Here $s$ is the arc length parameter and $\kappa(s)$ is
Euclidean (signed) curvature.  We consider the smooth plane curves
$\gamma:[0,\ell]\to\mathbb{E}^2$ with two endpoints and tangent vectors at the ends being fixed.
Then the Euler-Lagrange equation obtained from the variation of the curve is
\begin{equation}\label{Elastic}
2\kappa^{\prime\prime}+\kappa^3-\lambda\kappa=0,
\end{equation}
where $\lambda$ is a constant and $'=d/ds$ . For more information on the elasticae, we refer to
\cite{BG,Mumford}.

On the other hand, we consider the isoperimetric deformation of plane curves of the form
(\textit{Goldstein-Petrich flow} \cite{GP}, see also \cite{Lamb}):
\begin{displaymath}
\dot\gamma=-\kappa^\prime N^{\mathsf E}
-\frac{1}{2}\kappa^2 T^{\mathsf E}. 
\end{displaymath}
Here $\dot{}=\partial/\partial t$, $T^{\mathsf E}=\gamma^\prime$ and $N^{\mathsf E}=JT^{\mathsf E}$,
where $J$ is the positive $\pi/2$-rotation, are the unit tangent tangent vector field and the unit
normal vector field.  The Frenet frame $F^{\mathsf E}=(T^{\mathsf E},N^{\mathsf E})\in\text{SO}(2)$
satisfies
\begin{equation}
(F^{\mathsf E})^{-1}F^{\mathsf E}{}^\prime = \kappa J,\quad
(F^{\mathsf E})^{-1}\dot F^{\mathsf E}
=-\left(\kappa^{\prime\prime}+\frac{1}{2}\kappa^{3}
\right)\,J, \label{eqn:Lax_mKdV}
\end{equation}
where the first equation is the Frenet formula. 
The compatibility condition of \eqref{eqn:Lax_mKdV} yields the mKdV equation
\begin{equation}\label{eqn:mKdV}
\dot\kappa + \kappa^{\prime\prime\prime}+\frac{3}{2}\kappa^2\kappa^\prime=0. 
\end{equation}
In the terminology of integrable systems, \eqref{eqn:Lax_mKdV} is nothing but a {\em Lax pair} (for
vanishing spectral parameter).  The travelling wave solutions $\kappa(s-\lambda{t}/2)$ to the mKdV
equation satisfy $2\kappa^{\prime\prime} + \kappa^3 - \lambda\kappa=c_0$ for some constant
$c_0\in\mathbb{R}$.  Comparing with \eqref{Elastic}, we see that every elastica is a solution to the
Goldstein-Petrich flow.
%
\section{Log-aesthetic curves and similarity geometry}
\begin{figure}[t]
\centering
\includegraphics[width=0.4\linewidth]{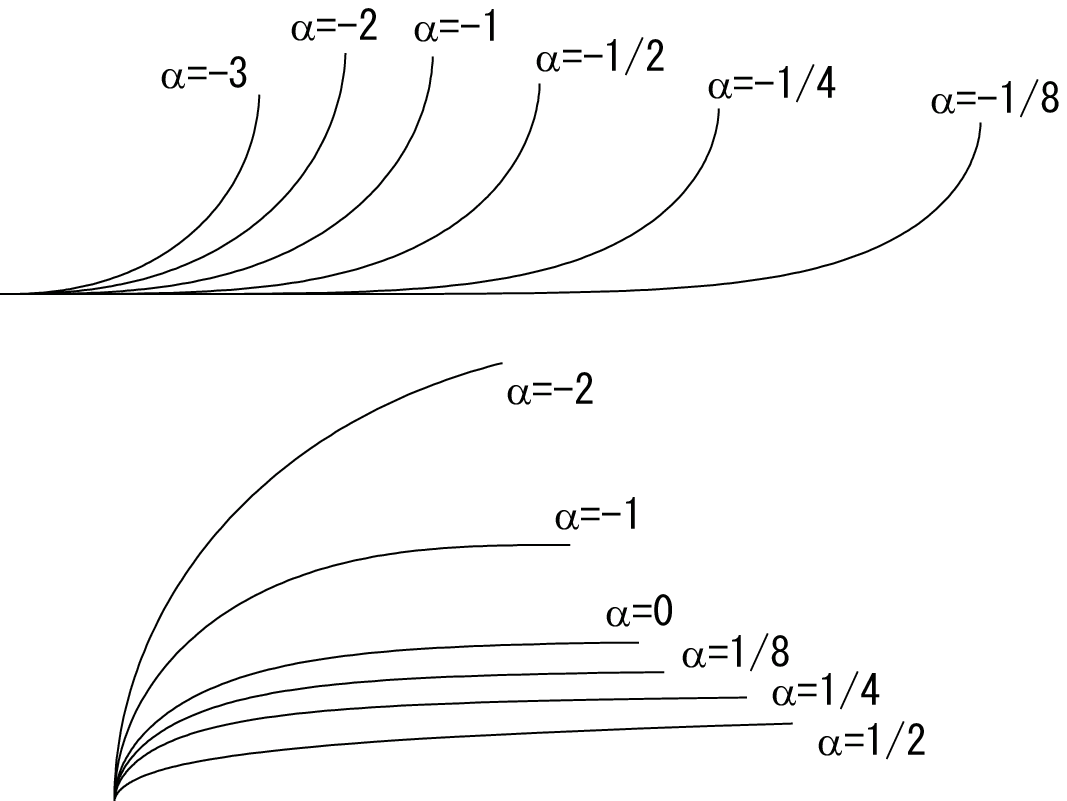}\qquad \includegraphics[width=0.25\linewidth]{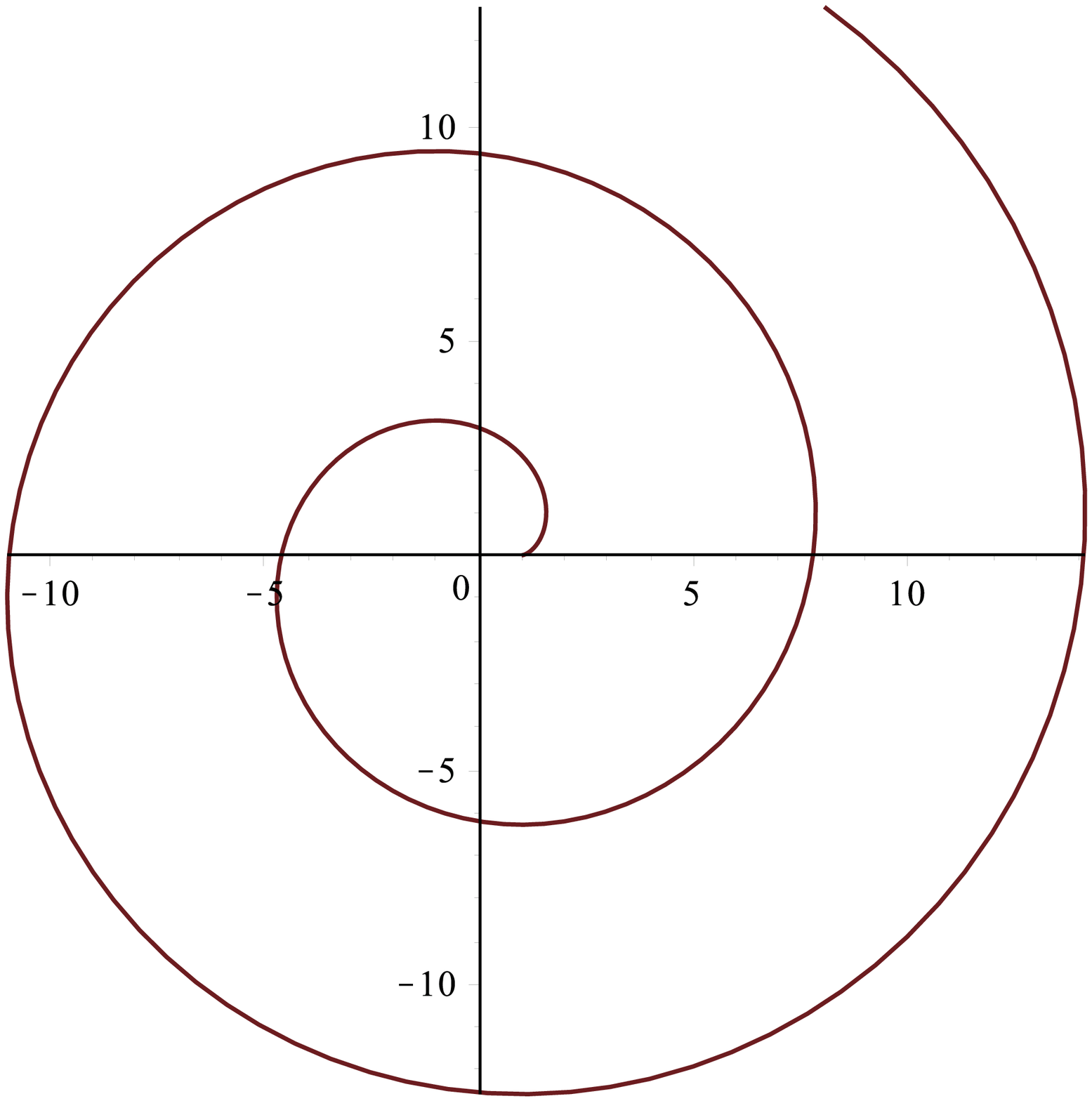}
\caption{Log-aesthetic curves. Left: LAC with various parameters. Right: LAC with $\alpha=2$
(circle involute curve).}\label{fig:LAC}
\end{figure}

According to \cite{Miura}, an arc length parametrized plane curve $\gamma(s)\in\mathbb{R}^2$ is said
to be an LAC of slope $\alpha$ if its signed curvature
radius $q$ satisfies
\begin{equation} \label{eqn:LAC_curvature}
 q(s)^{\alpha}= as+b\ \ (\alpha\not=0),\quad q(s)=\exp(as+b)
\ \ (\alpha=0),\quad a, b\in\mathbb{R}.
\end{equation}
The class of LAC's includes some
well known plane curves, for instance, the logarithmic spiral ($\alpha=1$), the clothoid
($\alpha=-1$) and the \textit{Nielsen spiral} ($\alpha=0$). Figure \ref{fig:LAC} illustrates
some examples of LAC.

Recent advancements on the LAC have been promising as indicated by Levien and S\'equin
\cite{Levien2009}. Theoretical backgrounds have been set up, such as 
variational formulation of LAC for free-form surface design \cite{Miura2012}, and fast
computational algorithm \cite{Ziatdinov2012}.  An important development for industrial design
practices has been presented in \cite{Meek2012}, where it is proved that a unique solution exists
for the $G^1$ interpolation problem using an LAC segment when $\alpha<0$. 
%
For more details of the LAC we refer to \cite{Miura2014,Miura2016}.
%
%
%

Those developments have been made based on the basic characterization \eqref{eqn:LAC_curvature} in
the framework of Euclidean geometry. However, \eqref{eqn:LAC_curvature} is too simple to identify
the underlying geometric structure. Consequently, we do not have a ``good'' guideline to generate
a larger class of ``aesthetic'' geometric objects including LAC based on a sound mathematical
background. In this paper, we clarify that LAC fits well the framework of the similarity geometry.

The similarity plane geometry is a Klein plane geometry associated with the group of similarity
transformations, \textit{i.e.}, isometries and scalings: $\mathbb{R}^2\ni \bm{p}\longmapsto
rA\bm{p}+\bm{b}$, $A\in{\rm SO}(2)$, $r\in\mathbb{R}_{+}$, $\bm{b}\in\mathbb{R}^2$.  The natural
parameter of plane curves in similarity geometry is the \textit{turning angle} $\theta=\int
\kappa(s)\>ds$.  Let $\gamma(\theta)$ be a plane curve in similarity geometry parametrized by
$\theta$.  The \textit{similarity tangent vector field} $T$ and the \textit{similarity normal vector
field} $N$ are defined by $T=\gamma^\prime$, $N=J\gamma^\prime$, respectively, where
$'=d/d\theta$. Then $T$ and $N$ satisfy the \textit{similarity Frenet equations}:
\begin{equation}
T^\prime=-uT+N,\quad N^\prime=-T-uN, 
\end{equation}
for some function $u(\theta)$ which is called the \emph{similarity curvature}. One can check that
plane curves in similarity geometry are uniquely determined by the similarity curvature up to 
similarity transformations.  The similarity curvature $u$ is related to the signed curvature radius
$q=1/\kappa$ by the \textit{Cole-Hopf transformation}:
\begin{equation}
u=-\frac{q'}{q}. \label{eqn:C-H} 
\end{equation}

Surprisingly, the notion of LAC is shown to be invariant under similarity transformations. For
instance, the slope $\alpha$ is expressed as $\alpha = 1+u^\prime /u^2$. LAC and qAC are reformulated in
terms of similarity geometry as follows \cite{Inoguchi, SS1, SS2}: A plane curve $\gamma(\theta)$ is
said to be a LAC of slope $\alpha$ if its similarity curvature satisfies the Bernoulli equation:
\begin{equation}\label{LAC}
u^\prime=(\alpha-1)u^2.
\end{equation}
%
Similarly, a plane curve is said to be a qAC of slope $\alpha$ if its
similarity curvature obeys the Riccati equation:
\begin{equation}
u^{\prime}=(\alpha-1)u^2+c,\quad c\in\mathbb{R}:\text{const}.
\label{eqn:Riccati_LAC}
\end{equation}
%
\section{Burgers flows on similarity plane curves}
It is known that plane curves in similarity geometry admit isogonal deformations governed
by the Burgers hierarchy \cite{CQ}.  The simplest flow is given by
\begin{equation}\label{CQflow}
\dot\gamma=(b-u)T-N,\ \ b\in\mathbb{R}:\text{const.}
\end{equation}
The similarity Frenet frame $F=(T,N)$ satisfies the similarity Frenet equation and the deformation
equation
\begin{equation}
F^{-1}F' =-uI+J,\quad 
F^{-1}\dot F = (-u'+u^2+1-bu)I + bJ, \label{eqn:similarity_frame}
\end{equation}
where $I$ is the identity matrix. The compatibility condition of \eqref{eqn:similarity_frame} yields
the \textit{Burgers equation}:
\begin{equation}
\dot u = (u'-u^2+bu)'. \label{eqn:Burgers}
\end{equation}
The Burgers equation is linearized in terms of the signed curvature radius via the
Cole-Hopf transformation \eqref{eqn:C-H} as
$\dot q = q^{\prime\prime}+bq' $.
We note that the parameter $b$ corresponds to reparametrization of the curve.
Imposing the \textit{stationary ansatz} $\dot u=0$, the Burgers equation \eqref{eqn:Burgers} reduces to the Riccati equation:
\begin{equation}\label{St-Burgers}
u^\prime = u^2-bu+c,\ \ c\in\mathbb{R}.
\end{equation}
We recover the Riccati equation \eqref{eqn:Riccati_LAC} with $\alpha=2$ by putting $b=0$. Note that
\eqref{eqn:Riccati_LAC} is obtained formally from \eqref{St-Burgers} by replacing the variable as
$u\to (\alpha-1)u$. In this sense, qAC is characterized as the stationary curves of the Burgers
flow.
%
\section{Fairing energy in similarity geometry}
Let us discuss the variational formulation of LAC and qAC.
We introduce the \textit{fairing energy functional}
$\mathcal{F}^{\lambda,a}$ by
\begin{equation}\label{eqn:fairing_functional}
\mathcal{F}^{\lambda,a}(\gamma)=\int_{\theta_1}^{\theta_2}
\frac{1}{2}
\left\{
a^2 u(\theta)^2+\lambda\left(\frac{q_1\,q_2}{q(\theta)^{2}}\right)^a
\right\}
\,\mathrm{d}\theta,
\end{equation}
where $a=\alpha-1$, $\lambda$ is an arbitrary constant and $q_i = q(\theta_i)$ ($i=1,2$).  The above
functional is invariant under similarity transformations and its name ``fairing energy'' is motivated
from the fairing procedure in digital style design of industrial products. We write the variation of
$\gamma$ as $\delta\gamma=\xi(\theta)T(\theta)+\eta(\theta)N(\theta)$.  Standard variational
calculus yields $\delta q/q = \phi - \psi^\prime$, $\delta u/u = -(u\psi^\prime +
\phi-\psi^\prime)$ and $\delta \theta = \psi$, where $\phi(\theta) = \xi' - \xi u - \eta$,
$\psi(\theta)=\eta' - \eta u + \xi$.  By using those data, we obtain the \textit{first variational
formula} of the fairing energy:
\begin{equation}\label{eqn:var_fairing_functional}
\delta\mathcal{F}^{\lambda,a}(\gamma)
= - \frac{1}{2}\left[ a^2u(\widetilde\phi-\widetilde\psi^\prime) + H(\gamma)\widetilde\psi\,\right]_{\theta_1}^{\theta_2} + 
\frac{a}{2}\int_{\theta_1}^{\theta_2}\left\{au'-\lambda\left(\frac{q_1q_2}{q^2}\right)^a\right\}
(\widetilde\phi-\widetilde\psi^\prime+u\widetilde\psi)\,{\rm d}\theta, 
\end{equation}
where $ \widetilde\phi = \phi - \frac{\phi(\theta_1)+\phi(\theta_2)}{2}$, $\widetilde\psi = \psi -
\frac{\psi(\theta_1)+\psi(\theta_2)}{2}$, and $H(\gamma) = a^2 u(\theta)^2
-\lambda\left(\frac{q_1\,q_2}{q^{2}}\right)^a$.  The first variation formula implies that if
$\gamma$ is a critical point of the fairing energy for deformations which respect the boundary
condition in \eqref{eqn:var_fairing_functional}, then $\gamma$ satisfies
\begin{equation}\label{eqn:st-Burgers2}
 au'-\lambda\left(\frac{q_1q_2}{q^2}\right)^a=0,
\end{equation}
which is equivalent to the Riccati equation for qAC \eqref{eqn:Riccati_LAC} together with
\eqref{eqn:C-H}.  Note that $H(\gamma)$ is a first integral of \eqref{eqn:st-Burgers2}.  If we
require preservation of the total similarity arc length $\delta(\theta_2-\theta_1)=0$, then 
the boundary term vanishes iff $\phi(\theta_1)=\phi(\theta_2)$. This implies that
$\frac{q_2}{q_1}$, namely, the ratio of length of tangent vectors at the endpoints is preserved
by the variation. Note that this condition is invariant with respect to similarity
transformations.
%
\begin{thm}\label{mainresult}
If a plane curve $\gamma$ is a critical point of the fairing energy $\mathcal{F}^{\lambda,a}$
\eqref{eqn:fairing_functional} under the assumption of preservation of total similarity arc length
and the boundary condition that the ratio of length of tangent vectors at the endpoints is
preserved, then the similarity curvature $u$ satisfies $u'=au^2+c$,
where $c$ is a constant. Therefore, quasi aesthetic curves of slope $\alpha\not=1$ are critical
points of the fairing functional.
\end{thm}
%
\begin{rem}{\rm
Moreton and S{\'e}quin \cite{MS} considered the functional
$\int_{\gamma}\left(\frac{\mathrm{d}\kappa}{\mathrm{d}s} \right)^2\>\mathrm{d}s\cdot \left(\int_{\gamma}\>\mathrm{d}s\right)^3$, 
which is invariant under similarity transformations.  A curve which is a critical point of this
functional is called a \emph{scale-invariant minimum variation curve} (SI-MVC) in \cite{BBK}.  This
is different from \eqref{eqn:fairing_functional}, since the above functional is rewritten as
$\left(\int\frac{u(\theta)^2}{q(\theta)^3}\>\mathrm{d}\theta\right)\>
\left(\int_{\theta_1}^{\theta_2}q(\theta)^3\>\mathrm{d}\theta\right)$.  }
\end{rem}
%
The mathematical characterization of LAC and qAC presented in this paper may enable us to generalize
the notion of aesthetic curves. For instance, it may be possible to construct discrete LAC and qAC
preserving underlying integrable structure, which may be useful for a numerical approach.
Associated results will be reported in a forthcoming paper. It is noted that discrete isogonal
deformations of discrete plane curves governed by the discrete Burgers equation have been studied in
\cite{KKM}.
%
\section*{Acknowledgment}
This work was partially supported by JSPS KAKENHI, JP25289021, JP16H03941, JP16K13763, JP15K04834,
JP26630038, JST RISTEX Service Science, Solutions and Foundation Integrated Research Program, ImPACT
Program of the Council for Science, Technology and Innovation. The authors acknowledge the support
by 2016 IMI Joint Use Program Short-term Joint Research No. 20160016. Y.~Shimizu acknowledges the
support by 2015 IMI Joint Use Program Short-term Visiting Researcher No.20150010. The authors would
like to express their sincere gratitude to Prof. Miyuki Koiso, Prof. Hiroyuki Ochiai, Prof. Nozomu
Matsuura and Prof. Sampei Hirose for invaluable comments and fruitful discussions.


\end{document}